\documentclass{emulateapj}
\shorttitle{GRB Correlations}
\shortauthors{Butler et al.}
\submitted{ApJ Accepted}

\def\gtrsim{\mathrel{\hbox{\rlap{\hbox{\lower4pt\hbox{$\sim$}}}\hbox{$>$}}}}
\def\lessim{\mathrel{\hbox{\rlap{\hbox{\lower4pt\hbox{$\sim$}}}\hbox{$<$}}}}
\newcommand{\beq}{\begin{equation}}
\newcommand{\eeq}{\end{equation}}

\begin{document}

\title{Generalized Tests for Selection Effects in GRB High-Energy Correlations}

\author{Nathaniel R. Butler\altaffilmark{1,2,3},
Daniel Kocevski\altaffilmark{2}, and Joshua S. Bloom\altaffilmark{2,4}.}
\altaffiltext{1}{Townes Fellow, Space Sciences Laboratory,
University of California, Berkeley, CA, 94720-7450, USA}
\altaffiltext{2}{Astronomy Department, University of California,
445 Campbell Hall, Berkeley, CA 94720-3411, USA}
\altaffiltext{3}{GLAST/Einstein Fellow}
\altaffiltext{4}{Sloan Research Fellow}

\begin{abstract}
Several correlations among parameters derived from modelling the high-energy 
properties of GRBs have been reported.  We
show that well-known examples of these have common features indicative of strong
contamination by selection effects.  We focus here on the impact
of detector threshold truncation on the spectral peak versus isotropic equivalent
energy release ($E_{\rm pk}$-$E_{\rm iso}$) relation, extended to a large sample
of 218 Swift and 56 HETE-2 GRBs with and without measured redshift.
The existence of faint Swift events missing from
pre-Swift surveys calls into
question inferences based on pre-Swift surveys which must be subject to
complicated incompleteness effects.  We demonstrate a generalized method for treating data
truncation in correlation analyses and apply this method to Swift
and pre-Swift data.
Also, we show that the $E_{\rm pk}$-$E_{\gamma}$ (``Ghirlanda'')
correlation is effectively independent of the GRB redshifts, which suggests
its existence has little to do with intrinsic physics.  
We suggest that a physically-based correlation, manifest observationally,
must show significantly reduced scatter
in the rest frame relative to the observer frame and must not persist
if the assumed redshifts are scattered.
As with the $E_{\rm pk}$-$E_{\gamma}$ correlation, we find that the pre-Swift, bright 
GRB $E_{\rm pk}$-$E_{\rm iso}$ 
correlation of \citet{amati06} does not rigorously satisfy these conditions.
\end{abstract}


\maketitle

\section{Introduction}
\label{sec:intro}

Correlations are pervasive in astronomy and generally lead theory in
allowing us to discover causal relationships between observed quantities.
In the study of Gamma-ray Bursts (GRBs), multiple powerlaw relations 
among GRB observables have been used to uncover the intrinsic physics
of GRBs themselves \citep[e.g.,][]{el04,yama04,rm05,le05,ldg05} and to 
use GRBs as probes to the distant Universe \citep[e.g.,][]{ghirl05,schaf07}.
Several authors have critically examined the limitations of
such relations: understanding how well they potentially constrain physics 
given their form and scatter \citep[e.g.,][]{fb05,schafCol07},
uncovering possible evolution with cosmic time \citep[e.g.,][]{yon04,li2007}, realizing
the commonality of outliers \citep{np05,bp05,kaneko06}, and --- most fundamentally ---
determining how the existence and form of the relations vary once spurious correlation
imparted by selection
effects is treated \citep[e.g.,][]{lpm00}.

Recently, \citet[][hereafter B07]{butler07} report evidence from Swift 
satellite \citep{gehr04} observations
of GRBs that several of the correlations
exhibit a wide scatter and shift in normalization
toward the Swift detection threshold, suggestive of an origin intimately connected to the
detection limits of pre-Swift satellites.
Because relations intrinsic to the physical processes underlying GRBs should not be
instrument-dependent, a broader investigation into the data from pre-Swift satellites
is crucial for determining whether and/or how the relations
can be trusted to potentially constrain the physics of GRBs or cosmology.

Here, we extend our critique to include the full B07 catalog of Swift GRBs with and
without measured
redshifts $z$ (for 218 total Swift GRBs) and also to include HETE-2 GRBs with and without
measured $z$ (56 GRBs).  
The uniform B07 catalog is novel for deriving bolometric fluences for all Swift
GRBs (between GRBs~041220 and 070509), without requiring tight error bars on the spectroscopic
fit parameters.  The sample is therefore flux limited by the sensitivity of
the Burst Alert Telescope \citep[BAT;][]{bart05}; because the X-ray Telescope \citep[XRT;][]{burrows05} localizes
to few arcsecond precision nearly all BAT GRB afterglows, additional flux limits associated
with afterglow localization and host galaxy detection (important for example when
considering the GRBs with measured $z$) are likely not present
or are far less important as compared to stronger flux limits imposed in pre-Swift
surveys.

A study similar to this has recently been conducted by \citet{ghirl08}
comparing a smaller sample of bright Swift GRBs to the Beppo-SAX sample (although only the sub-samples with
measured $z$).  \citet{ghirl08} find that faint GRBs detected by Swift
(and HETE-2; Section \ref{sec:hete})
are missing from the Beppo-SAX sample, and there is no clear explanation for the missing data
in terms of either the SAX trigger threshold or a (likely higher) threshold resulting from a
demand that tight error bars be derived in the GRB spectral modelling.
As in \citet{ghirl08}, we also focus
on the correlation between the
isotropic equivalent energy $E_{\rm iso}$ and the peak in the $\nu F_{\nu}$ spectrum
$E_{\rm pk}$ \citep{lpm00,amati02}.

Selection effects in high-energy GRBs observables are well-documented \citep[e.g.,][]{lp96}, if rarely treated.
For faint GRBs, an expected departure of the true distributions of flux, duration,
etc., from the observed distributions is expected given the
steep GRB number density versus peak photon flux $F$ relation \citep[$dN/dF \propto F^{-a}$, $a\sim 2$; e.g.,][]{preece00},
which places most GRBs in a given sample near the detection limit --- or near some other imposed
sample cutoff at a higher flux level --- with a narrow logarithmic dispersion of
$\ln(10)/(1-a)\sim 0.4$ dex.  

In Section \ref{sec:hete} of this paper we study the correlation of $E_{\rm pk,obs}$ with the energy fluence
$S \propto F E_{\rm pk,obs} dT$ roughly, where $dT$ is the burst duration, for which it is clearly important to compensate for
the expected non-intrinsic correlation of fluence with $E_{\rm pk,obs}$
that arises essentially because GRB detectors are photon counters and not bolometers.
Similarly structured correlations are then discussed in Section \ref{sec:anatomy}.

Two basic approaches have been attempted to compensate for GRB flux limits: (1)
forward folding of a model GRB rate density, evolution, and luminosity function
through a model of the detector response \citep[e.g.,][]{gdl05,ldg05}, and (2) applying non-parametric statistical methods to
the data and the observed flux limit for each data point.  In Section \ref{sec:hete}, we review and 
apply numerical methods which treat the observed data truncation directly according to (2),
while we plan to approach the B07 data via path (1) in a future paper.
In order to potentially better understand the origin of pre-Swift
correlations not strongly present in the Swift sample, we explore
the significance of these correlations after rejecting GRBs from the
Swift sample which would not have been detected by HETE-2 
(Section \ref{sec:hete}).

Extending our critique to the \citet{ghirl08} and
similar
pre-Swift samples of bright GRBs with redshifts \citep[e.g.,][]{amati02,amati06,schaf07} --- where the data truncations are poorly known
 and the GRBs come from multiple instruments of varying sensitivity --- we study (Section \ref{sec:anatomy}) 
how the relations transform from the observer to the source frames.  We show that
redshift dependence in the correlations is generally weak, which is expected for
a correlation that arises due to detection selection effects but also continues to have low scatter
in the source frame.  We present tests to uncover
whether the redshift dependence in other correlations is similarly weak,
possibly providing circumstantial evidence that these correlations are 
non-intrinsic.  This general approach, despite its limitations,
can potentially find application 
in other population analyses where thresholds are important.

\section{Flux Limits and the $E_{\rm pk}$-$E_{\rm iso}$ Correlation}
\label{sec:hete}

\begin{figure}
\centerline{\includegraphics[width=3.6in]{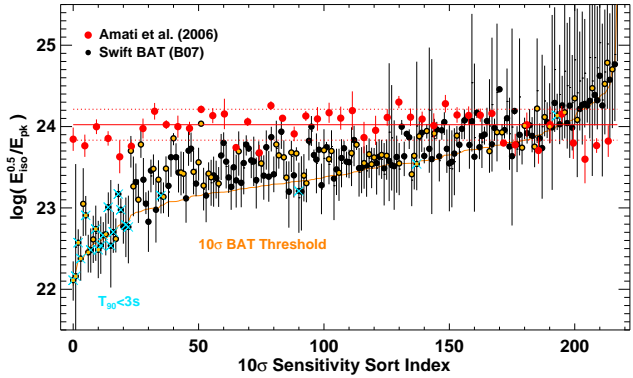}}
\caption{\small
The majority ($>67$\%) of Swift GRBs (black circles)
prefer a lower $E_{\rm iso}$-$E_{\rm pk}$ normalization
relative to the pre-Swift GRBs in \citet{amati06} (red circles).   
The Swift GRBs are sorted and indexed according to their $10\sigma$ limiting $E_{\rm iso}^{0.5}/E_{\rm pk}$ values
so that the GRBs detectable over the broadest range in the $y$ coordinate are plotted toward the left
of the plot.  Swift GRBs
with measured $z$ are marked with a small yellow circle.  For GRBs without measured $z$,
we assume $z=3.83$, which maximizes $E_{\rm iso}^{0.5}/E_{\rm pk}$ \citep[see, also,][]{np05}.
Swift short GRBs ($T_{90}<3$ s) are marked with a cyan $X$.
At these $z$, many (41\%) of the Swift sample GRBs exhibit $E_{\rm iso}^{0.5}/E_{\rm pk}$ values
inconsistent (at 90\% confidence) with the \citep{amati06} relation (red horizontal lines).
The \citet{amati06} consistent GRBs typically lie to the right of the plot, where $E_{\rm iso}^{0.5}/E_{\rm pk}$
sensitivity is weakest, indicating that
these GRBs would be undetectable at significantly lower fluence levels.}
\label{fig:ama}
\end{figure}

In Figure \ref{fig:ama}, we reproduce the narrow ($<0.2$ dex) scatter in $E_{\rm pk} \sim E_{\rm iso}^{0.5}$
observed in the recent review by \citet{amati06} for data from multiple 
missions.  Following \citet{np05} 
in realizing that $E_{\rm iso}^{0.5}/E_{\rm pk}$
reaches a maximum for $z\approx 3.83$, we assume this redshift 
value for all GRBs 
without measured redshift to place the entire B07 sample on the plot
(black points).
Most (67\%) best-fit Swift values are below the lower $1\sigma$ red-dotted \citet{amati06} line.
Nearly half (41\%) are below the red dotted line at 90\% confidence, indicating a clear preference for a lower flux normalization.
We determine 90\% confidence error bars directly for $E_{\rm iso}^{0.5}/E_{\rm pk}$, given the observed spectrum,
rather than attempting to propagate errors on the covariant quantities $E_{\rm iso}$ and $E_{\rm pk}$.

If we plot the BAT threshold corresponding to the best-fit points, we can see that agreement (right side
of the plot) is partly dictated by low-sensitivity, while many sensitive observations are strongly
inconsistent.  This finding is in excellent agreement with the BATSE studies of faint GRBs \citep{np05,bp05}.
There is no clear separation in Figure \ref{fig:ama} between 
the \citet{amati06}-consistent and \citet{amati06}-inconsistent
points, precluding obvious (non-circular) cuts which could create the 
semblance of close consistency.

In order to determine the most accurate threshold for each burst, we utilize the observed GRB spectrum and time profile 
directly instead
of employing a model for the threshold \citep[e.g.,][]{band03} which would only utilize results from the
spectral fitting (i.e., peak photon flux, $E_{\rm pk,obs}$, etc.) and possibly the burst duration \citep[e.g.,][]{band06}.
After determining the energy band --- taken to be 15-350 keV for Swift but allowed to vary for HETE-2 (see Figure \ref{fig:epsbol}) --- and 
temporal region which maximizes the signal-to-noise ratio $S/N$,
we divide the observed fluence by $(S/N)/10$.
This assumes a background dominated light curve (i.e., one where $\gtrsim 10$ times more counts reach the detector from the diffuse sky 
background and potentially from non-imaged sources not of interest
than from the source of interest), which is appropriate for all
but the brightest handful of Swift GRBs, for which our threshold calculation takes into account the observed background counts.

Here and below, we choose a cutoff $S/N=10$, because this is where the observed $S/N$ distributions appear to turn over.
The turn-over indicates a drop in detection
efficiency, because the number of faint bursts is known from BATSE \citep[e.g.,][]{preece00} to increase at low flux levels.
Because HETE-2 and Swift have intelligent trigger systems
which seek to find the light curve region maximizing the $S/N$ \citep[e.g.,][]{band06}, the counts that define the $S/N$ also
define the fluence threshold, approximately.  This threshold estimate is conservative
in cases where the trigger system fails to find the optimal $S/N$ region.

Many of the BAT points in Figure \ref{fig:ama} are likely lower than plotted, because most known GRBs have
$z<3.83$.  Also, the lower limit error bars in the figure for most of the points
are strongly influenced by a BATSE-based prior on $E_{\rm pk,obs}$, which introduces a bias 
against the rare but not-unprecedented (e.g. from BATSE) inference that $E_{\rm pk,obs}>300$ keV (see, B07).

Studies including some bright BAT GRBs \citep[e.g.,][]{amati06,schaf07} do not report the strong levels of inconsistency because
the studies exclude the faint or hard GRBs for which tight $E_{\rm pk,obs}$ constraints cannot be obtained.
However, for the simple question of consistency/inconsistency, tight $E_{\rm pk,obs}$ error bars are not required.

It is interesting to note that short-duration GRBs tend to appear in the lower left of Figure \ref{fig:ama}, populating
a region where GRBs of longer duration generally cannot be detected.  Only one short-duration GRB is detected at a flux
level consistent with the pre-Swift $E_{\rm iso}^{0.5}/E_{\rm pk}$ level.  As for long-duration GRBs,
short-duration GRBs tend to be detected near threshold. Appart from whether
$E_{\rm pk}$ correlates with $E_{\rm iso}$, because of the impulsive 
nature and resulting low fluence over peak flux
ratio for short-duration GRBs, short-duration GRBs are more likely than 
long-duration GRBs
to be outliers to an $E_{\rm pk}$-$E_{\rm iso}$ relation.  We caution, therefore, against using low $E_{\rm iso}^{0.5}/E_{\rm pk}$
values to classify \citep[as opposed to, e.g.,][]{amati08} intermediate duration events as short-duration (i.e., potential binary merger) events.

\subsection{From Swift BAT to HETE-2}

One straight-forward way to diagnose whether a flux limit is truly the origin of
inconsistency between the BAT and pre-Swift samples is to compare the BAT data to data from satellites
with very different detection sensitivities.  If the corresponding surveys both
extend to the respective limiting flux limits, then powerlaw fits for each sub-sample
should have appropriately different normalizations.
(In pre-Swift datasets considered alone as in the next section, the relative 
detection sensitivities generally do not vary widely,
and a different technique must be adopted.)  In this sub-section,
we compare the Swift and pre-Swift samples using a well-known approach which directly treats
the flux limits.

The application and refinement of statistical methods in survival analysis was
pioneered in BATSE GRB studies by \citet{vahe}, \citet{lp96}, and others.
By restricting comparison to ``associated sets'' of
data -- those GRBs detectable above the estimated thresholds of each
other GRB in the set -- it is possible to study distributions of and
correlations among observables independent of the thresholding, without making assumptions on
the nature of the GRBs below threshold.

In the early afterglow-era, these non-parametric methods were applied to correlations
between GRB source frame quantities \citep{lp99,lpm00,lr02,kl06}.   
The study of \citet{lpm00}, in particular, is important for
(1) demonstrating that the correlation between isotropic equivalent $\gamma$-ray
peak flux $L_{\rm iso}$ and $E_{\rm pk}$ likely
arises as a result of the detection process while also (2) discovering a possibly intrinsic
correlation between the isotropic equivalent energy release $E_{\rm iso}$ and $E_{\rm pk}$
\citep[also,][]{amati02}.  

The Kendall's $\tau_K$ statistic employed in these studies reports the fraction of concordant (i.e., correlating) data
minus the fraction of discordant data (i.e., anti-correlating).  Use of the statistic to rule on a
correlation's strength derives from maximum likelihood principles, and the $\tau_K$-test is non-parametric and
has maximum statistical power \citep{kend38}.
A simple, elegant, and rigorous extension of the $\tau_K$-test accounts for data truncation \citep[see,][]{efronpetrosion99}:
tabulation is restricted to the concordant or discordant data pairs which are detectable
above (or below in the case truncations from above) each other's limits.  In this way, both lower 
and upper limits on either the $x$ or $y$ variable can be treated, without the need to make assumptions regarding
the missing data.  Assumptions, sometimes difficult to uncover and possibly also hard to justify, regarding 
the missing data must generally be made with other methods \citep[e.g.,][]{gelman04}.

A major advantage of the truncated $\tau_K$-test (which employs the limits directly) is that it is in general much easier to
calculate a limit for an individual burst than it is to derive a general limit in $E_{\rm iso}$
valid for all $E_{\rm pk}$.  In fact, the general limit can never be calculated precisely, because the flux
limit (e.g., a satellite trigger threshold) is not a function of $E_{\rm iso}$, but of the observed photon flux in some
band pass over some shorter time integration.

\begin{figure}
\centerline{\includegraphics[width=3.6in]{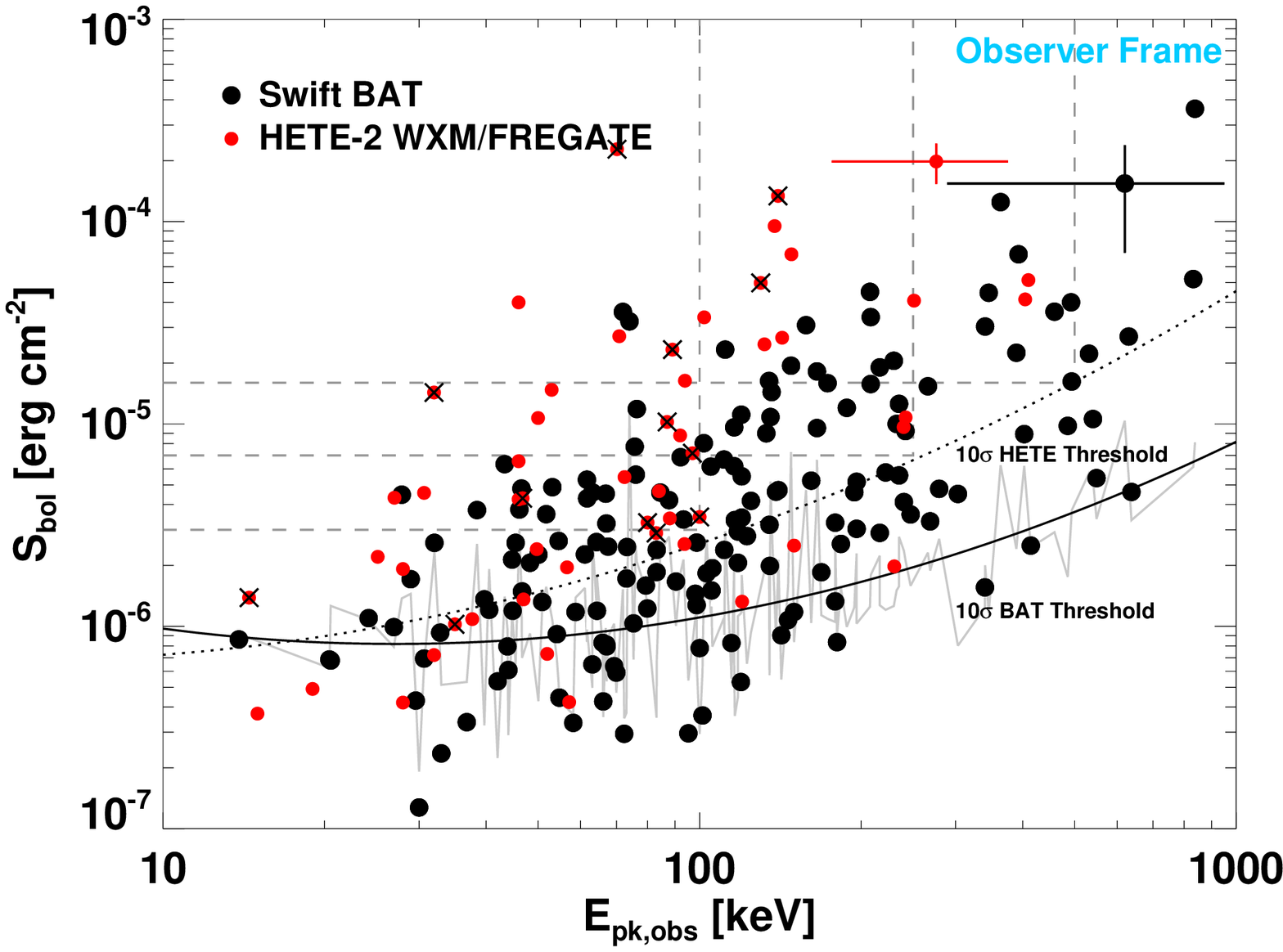}}
\centerline{\includegraphics[width=3.6in]{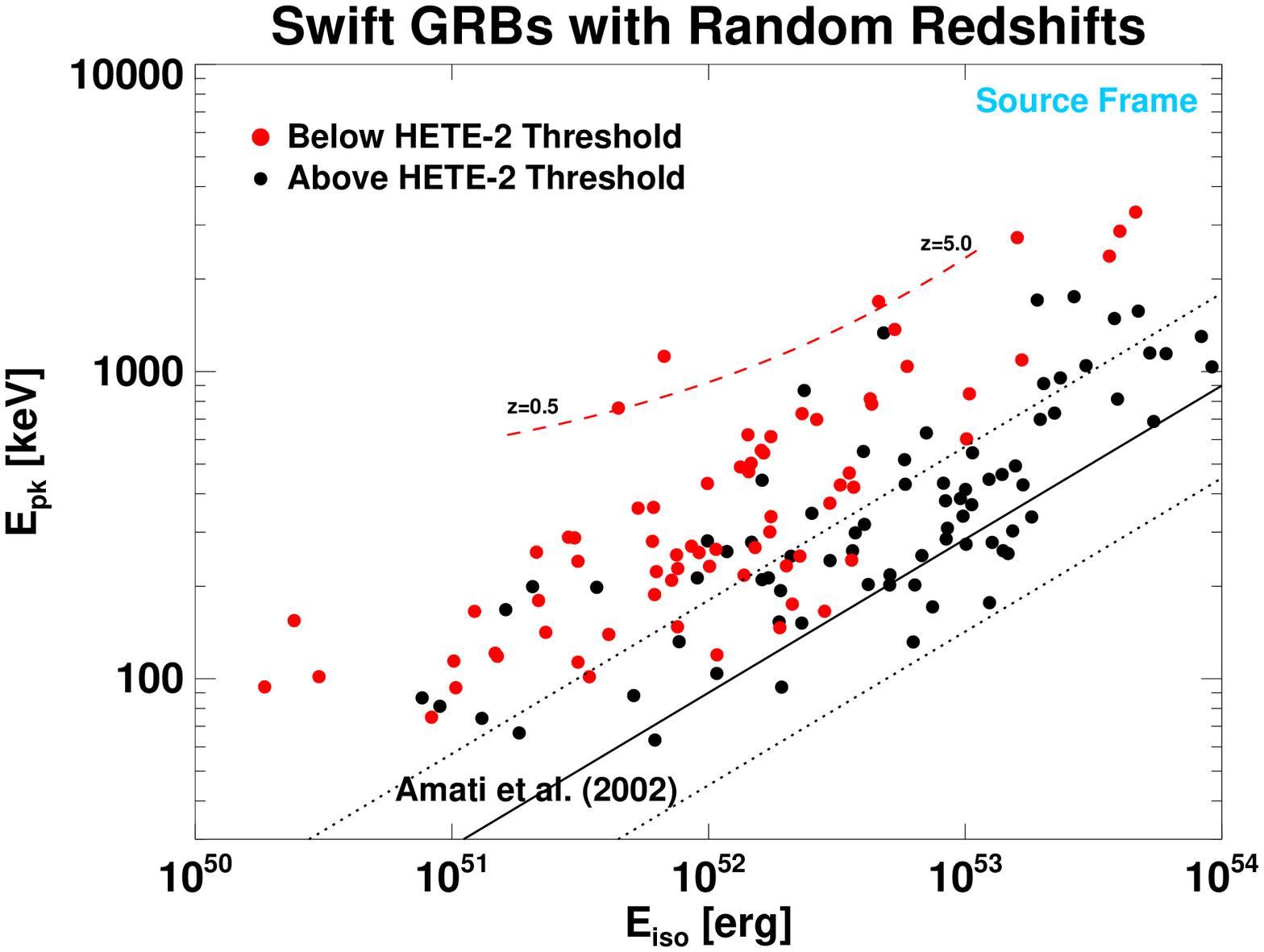}}
\caption{\small
From data truncation to spurious amplification of correlation strength in the observer and source
frames.  Typical error bars are plotted for points in the Top-Right corner.
(A) There is a strong ($>9\sigma$ significant) correlation in the observer
frame between $E_{\rm pk,obs}$ and $S_{\rm bol}$ for Swift (B07)
and HETE-2 \citep{taka}.  The jagged, gray line shows the observed
detection $S/N$ relative to $S/N=10$ for Swift GRBs, and the solid black line is
a fit to those values.  Similarly, using the larger $S/N$ value from 
either FREGATE
or WXM from the HETE-2 catalog \citep{roland}, we determine a $10\sigma$
threshold curve for HETE-2 bursts.  If we consider associated sets
\citep[see, e.g.,][]{vahe} of data
points above the HETE-2 threshold (e.g., dashed line regions), the
correlation in the Swift data is low ($2.4\sigma$).
The significance is still only $2.7\sigma$ if we lower the estimated HETE-2 threshold
by a factor of 2, an appropriate amount to account for error in our threshold
estimate (e.g., jagged line).
HETE-2 events with measured $z$'s are marked with an X.
(B) Randomly sampling $z$'s from Swift bursts, we show one iteration placing Swift points from
(A) onto the $E_{\rm pk}$-$E_{\rm iso}$ plot.  
 In $10^4$ iterations, only $(20\pm 3)$\% of Swift event below the HETE-2 threshold are consistent with the
 relation found by \citet{amati02}.}
\label{fig:epsbol}
\end{figure}

\begin{figure}
\centerline{\includegraphics[width=3.2in,height=1.8in]{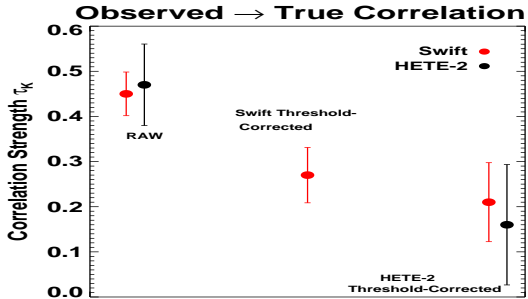}}
\caption{\small
The inferred $E_{\rm pk,obs}$-$S_{\rm bol}$ correlation strength for Swift and HETE-2, as measured with Kendall's $\tau_K$
coefficient, drops strongly as we correct for data lost to first the Swift threshold and then
the (higher) HETE-2 $S_{\rm bol}$ threshold (Section \ref{sec:hete}).
}
\label{fig:tau_plot}
\end{figure}

In the observer frame (Figure \ref{fig:epsbol}A), there is a highly 
significant apparent correlation between $S_{\rm bol}$ and $E_{\rm pk,obs}$
for both HETE-2 ($\tau_K=0.47$, $5.2\sigma$) and Swift ($\tau_K=0.45$, $9.3\sigma$);
see Figure \ref{fig:tau_plot}.  Here, $S_{\rm bol}$ is 
the energy fluence calculated in the source frame $1$--$10^4$ keV band
if $z$ is known \citep[e.g.,][]{amati02} or in the observer frame $1$--$10^4$ keV band if $z$ is unknown.
The significance of the correlation becomes modest ($\tau_K=0.27$, $4.4\sigma$) when we account
for the Swift threshold.  

That some significance remains indicates some amount of true correlation between $E_{\rm pk,obs}$ and $S_{\rm bol}$;
although, this residual correlation is strongly affected by the flux limit.
Quantifying the slope, scatter, etc., of this possible residual correlation and its origin in local intrinsic or population-wide
evolution effects will be the subject of a future study.  Here, we are interested in the expected appearance
of the correlation in the Swift survey relative to pre-Swift surveys.
The significance of the Swift ($\tau_K=0.21$, $2.4\sigma$) or HETE-2 ($\tau_K=0.16$, $1.2\sigma$)
correlation becomes marginal if we account for the HETE-2 detection threshold 
(see, Figure \ref{fig:epsbol}A).  To apply the HETE-2 threshold to Swift data, we fit a second order polynomial to the HETE-2 threshold and
evaluate that curve at the Swift $E_{\rm pk,obs}$ points.

In the source frame and restricting to the 63 normal, long-duration GRBs with $z$ in B07,
we find $\tau_K=0.59$ ($6.8\sigma$) for the $E_{\rm pk}$-$E_{\rm iso}$ relation.  However, considering only
associated sets of observations above Swift threshold, the correlation strength drops strongly ($\tau_K=0.28$, $2.4\sigma$).
The similarly strong decrease in correlation significance for the source and observer frame correlation can be understood by
noting that the correlations involve very similar data: observer frame points above and below the HETE-2 threshold transform 
to source frame points with a very weak dependence on the GRB redshifts.
The red dashed curve in Figure \ref{fig:epsbol}B ---
corresponding to the trajectory one event follows as its $z$
is varied --- illustrates how $z$ only weakly affects where
points fall on the plot (see also, Section \ref{sec:anatomy}).  
Points in the extreme Bottom-Right of Figure \ref{fig:epsbol}A  or Top-Left of Figure \ref{fig:epsbol}B 
cannot be made consistent at any $z$ \citep[Figure \ref{fig:ama}; see also,][]{np05,bp05}.

Randomly sampling redshifts from those observed by Swift, we find that $(80\pm 3)$\% of Swift 
events above the HETE-2 threshold are consistent with
the pre-Swift relation (solid and dotted lines).  Only $(20\pm 3)$\% of Swift event
below the HETE-2 threshold are consistent.  Most (60\%) of the events under
the HETE-2 threshold are inconsistent at any $z$.  We find similar results ---
$(62\pm 4)$\% consistency above HETE-2 threshold
and $(15\pm 3)$\% consistency below HETE-2 threshold --- if we assume a very different (toy)
$z$ distribution and draw $\ln(z)$ from a unit normal distribution.
The best-fit $E_{\rm pk}$-$E_{\rm iso}$ curve from B07 is a factor
$\approx 2$ higher than the pre-Swift curve (solid and dotted lines in Figure \ref{fig:epsbol}B).

\section{The Anatomy of a Correlation}
\label{sec:anatomy}

We now turn our attention to a set of correlation tests useful for
GRB samples with poorly understood flux limits.
As mentioned above, pre-Swift afterglow-era satellites do not have typical
sensitivities that vary by more than a factor of a few \citep[see also,][]{band03};
we find that diagnosing the reliability
of a correlation plotted on a log-log scale for these data cannot effectively be done by looking at variations in the
normalizations alone.  Moreover, the truncations tests discussed in the preceding section cannot easily be applied because
the flux limits for each GRB in most pre-Swift surveys are unpublished.
We seek to understand how the regression slope and scatter of the relation fits vary with assumptions on
the nature of the relations.

For simplicity, if we assume that all GRB fit 
parameters have the same error, the errors can be ignored.  Consider observer frame quantities
$X = \log{(S_{\rm bol})}$ and $Y=\log{(E_{\rm pk,obs})}$, for example.  These translate to source
frame quantities $X'= X + bZ = \log{(E_{\rm iso})}$ and $Y'=Y+aZ = \log{(E_{\rm pk})}$, for example.  Here,
$Z = \log{(1+z)}$ and $a=1$, $b\approx 3$ (accurate to 20\% for $0.5<z<5$, representing the bulk of known
GRB redshifts).

If we assume an origin in the observer frame, which we hope later to rule out, cross terms involving the observables and $z$ can be ignored,
and the expected source-frame linear regression slope \citep[see, e.g.][]{bevington} is:
\begin{equation}
m_{\rm src} = \left[ m_{\rm obs} + ab {\langle Z^2\rangle \over \langle X^2 \rangle} \right] / 
  \left[ 1+b^2 {\langle Z^2\rangle \over \langle X^2 \rangle} \right].
\label{eq:one}
\end{equation}
Here, $\langle .\rangle$ denotes an average after subtracting away the mean in $X$ or $Z$, and 
$m_{\rm obs}$ is the slope in the observer frame.  Equation \ref{eq:one} gives the slope of the {\it apparent}~source
frame relation given the observed data, whereas determination of the {\it true}~source frame relation requires detailed
knowledge of the GRB rate density and luminosity function to impute the missing data.
Although recent progress
has been made on measuring these \citep[e.g.,][]{kistler}, we do not attempt here to reconstruct $m_{\rm obs}[m_{\rm src,true}]$.
Moreover, for comparison with most GRB studies which simply assume $m_{\rm src,true}=m_{\rm src}$ \citep[e.g.,][]{firm04}, we only
require Equation \ref{eq:one}.

Relatively independent of which survey we use, 
$\langle Z^2\rangle / \langle X^2 \rangle \approx 0.1$, and $m_{\rm src} \approx m_{\rm obs}$.
This is the typical expected behavior for GRB correlations where $X$ or 
$Y$ are typically much
more broadly distributed than $Z$.  
There are, however, example correlations \citep[e.g., consider][]{willin07}
where the observables $X$ and $Y$ are arranged such that 
$m_{\rm obs}\approx 0$, and the scatter in $Z$ relative to that in
$X$ can be important for defining the chance $m_{\rm src}$ according to
Equation \ref{eq:one}.

We define the scatter $\sigma$ in $Y$ about the best-fit regression line to
be the root-mean-square (RMS) deviation.
For the best-fit regression line in Equation \ref{eq:one}, the expected
scatter is
\begin{equation}
\sigma^2_{\rm src} = \langle(Y-m_{\rm src}X)^2\rangle + (a-m_{\rm src}b)^2\langle Z^2 \rangle.
\label{eq:two}
\end{equation}
Covariance terms between $X$ and $Z$ or between $Y$ and $Z$ --- which generally 
act to decrease the scatter in the source frame as compared to the observer frame --- 
are dropped because we assume the correlation is non-intrinsic.  To be explicit, rearrangement of an equation like $E_{\rm pk} \propto
E_{\rm iso}^{0.5}$ shows that the observables should vary with $z$, and this implies cross terms in Equation \ref{eq:two}.
The cross terms go to zero if the correlation is due to ($z$-independent)
observational effects.

The first term in Equation \ref{eq:two} is bound from the bottom by $\sigma^2_{\rm obs}$; equality is
obtained for $m_{\rm src} = m_{\rm obs}$.  The second term in Equation \ref{eq:two} implies an increase in $\sigma^2_{\rm src}$
over $\sigma^2_{\rm obs}$ unless, coincidentally, $a=m_{\rm src}b$.  Interestingly, this ``coincidence'' is satisfied nearly
or exactly for many of the known GRB correlations.  

As an example, consider the \citet{ggl04} correlation: $E_{\rm pk} \propto E_{\rm \gamma}^{0.7}$,
where $E_{\rm \gamma} \propto E_{\rm iso} \theta_{\rm jet}^2$ is the energy release corrected for beaming
into a jet of angle $\theta_{\rm jet}$.  For the case of a uniform density medium surrounding the
GRB \citep{sari99}, $\theta_{\rm jet}^2 \propto E_{\rm iso}^{-1/4} (1+z)^{-3/4} T_{\rm jet}^{3/4}$, where $T_{\rm jet}$
is the time in the observer frame where the jetting effects are expected to become apparent.  Substituting into
equations \ref{eq:one} and \ref{eq:two}, it follows that $X=3/4\log{(S_{\rm bol}T_{\rm jet})}$, $Y=\log{(E_{\rm pk,obs})}$, $a=1$ 
and $b\approx 3/2$.  Hence, $a\approx m_{\rm src}b$,
as also demonstrated graphically in Figure \ref{fig:ghirl}A.  The \citet{ggl04} correlation is, therefore, effectively independent of the
measured spectroscopic redshifts. 

\subsection{Example Numerical Tests}

\begin{figure}
\centerline{\includegraphics[width=3.6in]{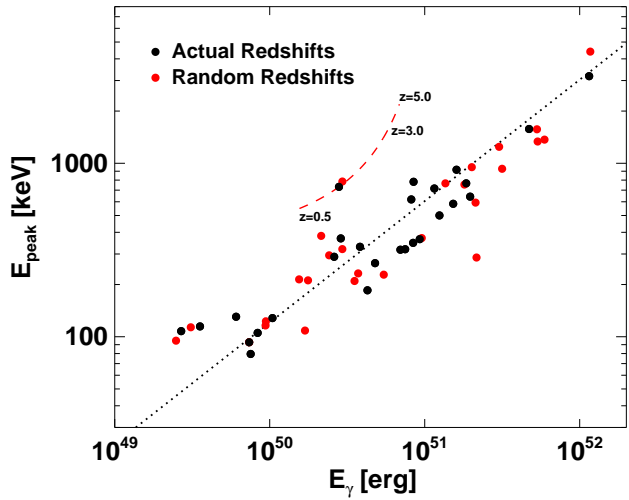}}
\caption{\small
\citet{ggl04} find a tight correlation between the beaming-corrected GRB
energy release $E_{\gamma}$ and $E_{\rm pk}$.  We also plot recent points
from \citet{schaf07}.  Suggesting a non-intrinsic origin similar to the
correlations in Figure \ref{fig:epsbol}, it is not necessary to know the
$z$ of an event to acceptably place it on the best-fit curve (dotted line).
The dashed red curve shows the trajectory with $z$, holding the other
observables fixed.
The tightness of the correlation is unaffected whether
true or random redshifts are assumed for translating the observer frame
to source frame quantities.}
\label{fig:ghirl}
\end{figure}

For the \citet{ggl04} correlation data in 
Figure \ref{fig:ghirl}A, the RMS scatter is $\sigma_{\rm obs}=0.17$ dex
for the observer frame variables
and $\sigma_{\rm src}=0.16$ dex for the variables corrected for 
cosmological distance and redshift.
This is a $0.9\sigma$ insignificant 
decrease according to an F-test.  Likewise, we find that the source frame 
data are correlated
strongly with Kendall's $\tau_K=0.7$.  However, data generated with 
randomized redshifts
 (e.g., red points in Figure \ref{fig:ghirl}A)
exhibit larger $\tau_K$ in most (76\%) of the simulations.  
Somewhat more encouraging, we find a marginal $2\sigma$ increase in significance
in $\sigma_{\rm src,sim}=0.18\pm 0.01$ dex.  However, the increase is
apparently dominated by 1--2 outlier events in a fraction of the simulations;
if we instead employ the median absolute deviation about the median of the fit
residuals as a robust measure of scatter, 52\% of simulations exhibit
lower scatter than for the observed data.  

For comparison with the previous section and considering
the 47 normal, long-duration bursts in \citet{amati06} for the 
$E_{\rm pk}$-$E_{\rm iso}$ relation, we find
a significant change in RMS scatter and slope only if we include two events at 
$E_{\rm iso}<2 \times 10^{51}$ erg.  
Using the outlier-resistant measures of correlation ($\tau_K$ or 
scatter estimated via the median absolute deviation about the median)
we find that 10\% of simulations yield a better correlation than the 
observed correlation, independent
of whether the two XRFs are excluded.  Hence, there is only very weak 
($1.5\sigma$) evidence from our tests to favor an intrinsic explanation for the $E_{\rm pk}$-$E_{\rm iso}$
correlation using the bright \citet{amati06} data.
Exclusion of a small number of outliers is common practice \citep[e.g.,][]{amati06}; however, it may be 
reasonable to criticize us
for allowing the outlier events to be flagged during and not before 
the simulations.

As an interesting side-note which may further suggest a paucity of intrinsic 
physics in \citet{ggl04} type relations, the balancing with $z$ is 
characteristic of similar relations found assuming instead a wind-stratified 
medium \citep{nava06} or even allowing $\theta_{\rm jet}$ to
vary arbitrarily with $T_{\rm jet}$ and $E_{\rm iso}$ \citep{liangzhang}.  
The latter relation has the smallest $\sigma_{\rm src}$, which leads
us to suspect that the only reason that a very large number of correlations 
do not abound in the literature is that we typically
restrict to those than can be interpreted physically, or, short of this, we 
require the observables to
evolve with $z$ in a physically plausible fashion 
(e.g., $T_{\rm jet,src} = T_{\rm jet}/[1+z]$).  

Without these constraints, it is possible 
to invent highly-significant but absurd relations (e.g., redshifted Swift trigger number versus redshifted burst duration)
 which make a mockery of all correlation studies.  
This effect alone is likely not sufficient to generate the tightness of the \citet{ggl04} correlation, however.

The tightness of the \cite{ggl04} correlation, if it is not intrinsic, must stem additionally from the effects outlined
in the previous sections for the functionally similar $E_{\rm pk}$-$E_{\rm iso}$ correlation.  
Excluding GRB~970508 as in \cite{ggl04}, we find that the $E_{\rm pk}$-$E_{\rm iso}$ relation has
only a slight increase in scatter (0.18 dex) relative to the $E_{\rm pk}$-$E_{\gamma}$ relation (0.16 dex), considering
the same data (Figure \ref{fig:ghirl}).
In fact, because the inferred
$T_{\rm jet}$ values are very narrowly distributed and $E_{\gamma}$ is proportional to $E_{\rm iso}$ to a power less than unity, 
a $E_{\rm pk}$-$E_{\gamma}$ relation is potentially always tighter than a $E_{\rm pk}$-$E_{\rm iso}$ relation, depending on
how the scatter is measured.  

\section{Discussion}

The B07 Swift BAT and \citet{taka} HETE-2 GRB samples both exhibit a statistically
significant correlation in the observer frame between $S_{\rm bol}$ and $E_{\rm pk,obs}$.  
What creates these correlations?  The
correlations follow the trigger threshold limits (Figure \ref{fig:epsbol}A) in both cases, and the correlation significances 
drop precipitously when we account for the flux limits (Section \ref{sec:hete}).  Therefore,
flux limits must play a strong role shaping the observed correlation, perhaps giving them most of their
statistical significance.

What does this tell us about the source frame $E_{\rm pk}$-$E_{\rm iso}$ relation?
Because the redshift dependence in the transformation from $E_{\rm pk,obs}$-$S_{\rm bol}$ to
$E_{\rm pk}$-$E_{\rm iso}$ is weak, the source frame correlation is likely to have the same
origin as the observer frame relation (Section \ref{sec:anatomy}).
Indeed, we observe a strong decrease in the source frame correlation significance when we account for
the Swift threshold.  

Finally, what can be learned about the $E_{\rm pk}$-$E_{\rm iso}$ relation as
it appears in Swift data relative to the pre-Swift $E_{\rm pk}$-$E_{\rm iso}$ relation?
The normalization appears to be strongly instrument-dependent, and this suggests
the pre-Swift normalization is defined largely by pre-Swift satellite flux limits.
B07 find that the Swift $E_{\rm pk}$-$E_{\rm iso}$ relation has a lower flux
normalization and more scatter relative to pre-Swift $E_{\rm pk}$-$E_{\rm iso}$ relations.
We show above that we can raise this normalization (and decrease the scatter about the relation) 
to a level consistent with the pre-Swift $E_{\rm pk}$-$E_{\rm iso}$ relation 
by imposing a heightened flux limit corresponding to detection by a satellite of HETE-2-like sensitivity.

Strictly speaking, from these observations we can only rule out
the existence of a narrow relation 
between $E_{\rm pk}$ and $E_{\rm iso}$, while some physical correlation 
between the quantities may be present at high flux levels (see, e.g., the 
paucity of bright events in Figure \ref{fig:epsbol}B).
In any case, future satellites more sensitive than Swift are expected to 
further shift and
broaden the $E_{\rm pk}$-$E_{\rm iso}$ relation into an inequality. 

More strongly, because we can probably attribute the slope (see, Section \ref{sec:anatomy}), scatter, and 
normalization of the correlation to selection effects related to a photon flux $F$ cutoff and the 
functional correlation between $E_{\rm pk,obs}$ and $S_{\rm bol} \propto F E_{\rm pk,obs} dT$
\citep[Section 1; also,][]{mass07}, an intrinsic explanation for the correlation in the Swift data may be unnecessary.
Our results also
call into question any intrinsic explanation for the correlation in pre-Swift data or in surveys
only including the brightest Swift events \citep[e.g.,][]{taka08}, because
an $E_{\rm pk}$-$E_{\rm iso}$ relation should be instrument-independent.
The existence of faint Swift events (in the observer frame) mostly missing from pre-Swift surveys shows that we
have a poor understanding of the observational selection effects which truncate the earlier data
\citep[see, also,][]{ghirl08}.  

We could be incorrect in drawing these conclusions if: (a) our spectral fits are systematically incorrect and
there are, in fact, no GRBs at flux levels below HETE-2 threshold levels, or (b) we have estimated the HETE-2
threshold incorrectly by a factor $>2$.  The accuracy of the spectral fits is addressed extensively in B07,
where direct consistency is established relative to observations from Konus-WIND or Suzaku of bursts also detected by Swift.
We note that \citet{bellm} have independently verified the statistical analysis in B07 for several bursts
also observed by RHESSI.  Assumptions in B07 regarding the GRB spectra (see, Section \ref{sec:hete}),
useful in compensating for the narrow BAT bandpass when determining bolometric fluences,
are conservative in that they err toward large $E_{\rm iso}^{0.5}/E_{\rm pk}$ values.
It also seems unlikely that the HETE-2 threshold should have any bearing
on the censoring of Swift GRBs, considering also that many authors have reported BATSE GRBs detected
below HETE-2 fluence limits \citep{np05,bp05,kaneko06}.  

For (b), we have shown that the $S_{\rm bol}$-$E_{\rm pk,obs}$ 
correlation significance, after accounting for the HETE-2 threshold, does not increase if we lower our threshold estimates
by a factor of two.  A larger error than this on our part for many GRBs is extremely unlikely given the straight-forward
nature of the threshold calculation (Section \ref{sec:hete}).  Moreover, such a substantial increase in HETE-2 sensitivity would make
HETE-2 as sensitive as Swift, which is demonstrably incorrect given the dramatic difference in the GRB localization rate of the
two missions ($\sim$20$/$yr for HETE-2 and $\sim$90$/$yr for Swift).  Given the mean peak photon flux to energy fluence
ratio from \citet[][$10^{8.0}$ $E_{\rm pk,obs}$ ph s$^{-1}$ erg$^{-1}$ keV]{taka}, our HETE-2 threshold 
estimate is within 50\% of that estimated in \citet{band03}.

Finally, there is strong indication that flux limits associated with spectroscopic redshift determination
are important to estimate and consider when restricting to GRBs with measured redshift \citep[e.g, as in][]{ghirl08}.
In Figure \ref{fig:epsbol} we mark the HETE-2 events
with measured redshift; these are on average a factor two brighter relative to threshold than the events
without measured redshifts.  

The quality of the HETE-2 GRB localization and possibly also the brightness of
the optical transient depend on the brightness of
the GRB; both effects contribute to
whether the afterglow and host galaxy can be detected.
Additional complicated selection effects related to determining the GRB $z$ are discussed
in \citet{bloom03}.  These are expected to be less important for Swift due to arcsecond X-ray localizations.

{\it We stress that a correlation is not de facto intrinsic
simply because trigger thresholding can be ruled out as influencing the surveys}.
Other flux limits may dominate (and probably do dominate when the samples are restricted to GRBs with measured
redshift).
A crucial step in utilizing pre-Swift data to rule on the nature of correlations is to
establish completeness in the surveys.

\subsection{New Correlations: Handle with Care}

To gauge the importance of systematic effects
in this area of research, we have isolated from the literature two paths that likely
have lead to apparently highly significant correlations:
(1) selection effects truncate the data in various ways and the ``missing'' data are not treated; or
(2) partial correlation with a hidden variable or variables is ignored.
In this paper, we have studied type (1) errors.
However, both type (1) and (2) errors can be unmasked using the tools outlined above.

Type (2) errors potentially arise in studies which employ one correlation \citep[e.g., Lag-Luminosity,][]{norris00} to infer $z$ for
  another correlation with similar variables (e.g., $E_{\rm pk}$-Luminosity), without controlling
  for partial correlation with the variable in common \citep[see, e.g.,][]{lr02,kl06}.

For example, from BATSE satellite observations where the detection threshold is very well characterized in terms of
peak luminosity,
we know that the $E_{\rm pk}$-Luminosity correlation is largely formed in the detection process \citep{lpm00}.
However, this potential strong bias is ignored when \citet{yon04} consider a purely intrinsic $E_{\rm pk}$-Luminosity to derive
a significant evolution in the rest-frame properties of GRBs (e.g., the correlation between Luminosity and $z$).
The potential instrumental origin of the $E_{\rm pk}$-Luminosity relation is also ignored by \citet{schaf07}.

We recommend testing potential new correlations (or known correlations not 
explicitly mentioned here) by proving:
\begin{enumerate}
\item Increase in RMS scatter $\sigma_{\rm obs}>\sigma_{\rm src}$ is 
statistically significant 
according to an F-test, and

\item the correlation scatter or significance determined using outlier 
resistant measures strongly decreases when the $z$'s are 
randomized and the correlations are recalculated, and

\item the observables, grouped to one side of the correlation equation,
vary with $z$ as predicted.

\end{enumerate}
These tests are to establish basic confidence in an intrinsic nature
for the correlations and are not merely to establish that a
correlation can be used to estimate $z$ \citep[e.g.,][]{li06,schafCol07}.

For these tests to be accurate, however, it is also
necessary to identify selection effects acting on the data.
Flux or fluence values should be present down to the established survey completeness level.
Well-established methods can then be applied to compensate for data truncation (Section \ref{sec:hete})
and to control for partial correlations \citep[e.g.,][]{as96}.
Covariance between the measured quantities, if present, must also be treated \citep[e.g.,][B07]{lp96,cabrera07}.

\section{Conclusions}

We show above for GRBs observed with multiple satellites 
that four example correlations reported in the literature
have features (instrument-dependent normalizations, weak $z$-dependence, etc.)
indicative of strong contamination by or even an origin in selection effects.
Contrarily, there is no widely-accepted, non-a-posterior explanation for 
the correlations in the source frame.
Also, the a-posteriori theoretical explanations \citep[e.g.,][]{el04,schaf07} fix the correlation normalizations
by requiring GRBs to have one intrinsic spectrum or to be
standardizable candles \citep[e.g., a narrow $E_{\gamma}$ distribution; see,][]{bfk03}, a hypothesis no longer 
well-supported by the data \citep[see,][]{koc08}.

The common $z$-independence of GRB correlations is either very odd or damning.  As we discuss, this is
one characteristic of a tight, apparent source-frame correlation which arises purely due to selection effects.
For correlations between luminosity and some other measured quantity that does not depend on luminosity distance, $z$-independence
is likely not a consequence of intrinsic physics, because the GRB cannot possibly know how the distance
to the observer should vary with $z$.  However, $z$-independence
might be expected for correlations between luminosities
\citep[e.g., the recently reviewed $E_{\rm iso}$-$L_{X,10}$ correlation,][]{nys08}.

We note that a requirement of $z$-independence trivially explains the 
relative slopes of 
the $E_{\rm pk}$-$E_{\rm iso}$ and
$E_{\rm pk}$-$E_{\gamma}$ relations (Section \ref{sec:anatomy}), something theory can apparently do as well but only with 
reference to complicated biases stemming from GRB beaming \citep{le05}.

As an obvious point, we caution that the balancing of 1$+$$z$ terms on
both sides of the
correlation equations (from which the approximate $z$-independence arises)
implies the correlations cannot be used to infer $z$.
More speculatively --- because redshift balancing allows a non-intrinsic observer-frame correlation to appear in the source frame as
a low-scatter correlation --- this balancing may have played a role in 
the discovery of potential intrinsic correlations.
If the correlations currently known have been selected in place of correlations that do not balance redshift,
and if the fitting of GRB properties and the pruning of outliers has been conducted with these correlations in mind,
then using these correlations to construct a Hubble diagram and test concordance cosmology \citep[e.g.,][]{ghirl05,schaf07} 
is circular.  

We stress that these redshift dependency problems --- and our intrepetation of them ---
are unique to and potentially only characteristic of high-$z$ objects like GRBs (as opposed to, e.g., SNe) where there is 
essentially no low-$z$ calibration,
and luminosity distance must generally be calculated using $z$ and assuming a cosmology model.

Small sample sizes may also have played an important role in allowing for tight apparent correlations
through over-fitting.  The correlation involving Luminosity, $E_{\rm pk}$, and $T_{45}$ duration for 22 GRBs
in \citet{firm06}, which does not appear to suffer from redshift balancing \citep[see Figure 7 in][]{firm06},
is an interesting case.  When looking at larger datasets, however, there appears to be increased scatter and
no statistically significant
improvement in scatter relative to the $E_{\rm pk}$-Luminosity \citep{colschaf08} or $E_{\rm pk}$-$E_{\rm iso}$
\citep{rossi08} correlations.  Investigation of the correlations in the largest possible datasets is,
therefore, critical.

The GRB community is no longer starved for data. The next critical step 
toward uncovering intrinsic correlations is to combine all available
data by establishing sample completeness in pre-Swift surveys \citep[e.g.,][]{ghirl08} and treating the dominant
flux truncations using methods like those outlined above.

In looking for new 
relations relevant to the physical processes underlying GRBs, and to avoid 
an inherent difficulty in
deciding an origin for the correlations in the source or observer frames (Section \ref{sec:anatomy}), it may be important to choose observables
less-broadly distributed than the characteristic range in 1$+$$z$ or to abandon observables with strong and complicated 
truncations (i.e., fluxes or fluences) altogether.  
Functional correlations in 1$+z$ could be minimized by choosing observables (e.g., powerlaw indices)
which do not vary explicitly with $z$.  Most directly and circumventing
all concerns raised above, we should establish calibration for GRBs 
at $z\lessim 0.1$.  It is, therefore,
also crucial to understand whether GRB properties evolve with redshift.  

\acknowledgments
N.R.B gratefully acknowledges support from a Townes Fellowship at the U.~C.  
Berkeley Space Sciences Laboratory, as well as partial support from J.S.B. 
and A. Filippenko, as well as support through the GLAST
Fellowship Program (NASA Cooperative Agreement: NNG06DO90A).
J.S.B. is partially supported by the Hellman Faculty Fund.

\end{document}